\documentclass[11pt, a4paper, titlepage]{article}
\usepackage{graphicx}
\usepackage{amssymb}

\newcommand{\bra}{\left\langle}
\newcommand{\ket}{\right\rangle}
\newcommand{\bral}[1] {\left \langle #1 \right |}
\newcommand{\lket}[1] {\left | #1 \right \rangle}
\newcommand{\braket}[2] {\left \langle #1 \right. \left | #2 \right \rangle}

\newcommand{\pder}[2]{\frac{\partial #1}{\partial  #2}}
\newcommand{\pdert}[2]{\frac{\partial^2 #1}{\partial  #2^2}}
\newcommand{\der}[2]{\frac{\mathrm{d} #1}{\mathrm{d}  #2}}
\newcommand{\dert}[2]{\frac{\mathrm{d}^2 #1}{\mathrm{d}  #2^2}}

\renewcommand{\d}{\mathrm{d}}
\newcommand{\e}{\mathrm{e}}

\newcommand{\ve}{\varepsilon}
\newcommand{\fp}{f_{\rm p}} 
\newcommand{\bfp}[1] {\bar f_{\mathrm{p}, #1}}
\newcommand{\hfp}[1] {\hat f_{\mathrm{p}, #1}}
\newcommand{\kb}{k_{\rm B}}

\newcommand{\LFP}{\mathcal{L}}
\newcommand{\Ps}{P^\mathrm{st}} 
\newcommand{\Pc}{P^\mathrm{c}}

\newcommand{\vtr}[2]{\left( \begin{array}{c} #1 \\ #2 \end{array} \right)}
\newcommand{\mtx}[4]{\left( \begin{array}{cc} #1 & #2 \\ #3 & #4 \end{array} \right)}

\begin{document}

\title{Fluctuations, Responses and Energetics of Molecular Motors}
\author{Takahiro Harada\footnote{E-mail: harada@life.ne.his.fukui-u.ac.jp} \\ {\it \small Department of Physics, Graduate School of Science, Kyoto University, Kyoto 606-8502, Japan\footnote{Current address: Department of Human and Artificial Intelligent Systems, University of Fukui, Fukui 910-8507, Japan}} \and Shin-ichi Sasa\footnote{E-mail: sasa@jiro.c.u-tokyo.ac.jp} \\ {\it \small Department of Pure and Applied Sciences, University of Tokyo, Komaba, Tokyo 153-8902, Japan}}
\date{Feburary 28, 2006}
\maketitle

{\bf Abstract}
\begin{quote}
A novel equality relating the rate of energy dissipation to a degree of violation of the fluctuation-response relation (FRR) in nonequilibrium Langevin systems is described. The FRR is a relation between the correlation function of the fluctuations and the response function of macroscopic variables. Although it has been established that the FRR holds in equilibrium, physical significance of violation of the FRR in nonequilibrium systems has been under debate. Recently, the authors have found that an extent of the FRR violation is related in a simple equality to the rate of energy dissipation into the environment in nonequilibrium Langevin systems. In this paper, we fully explain the FRR, the FRR violation, and the new equality with regard to a Langevin model termed a Brownian motor model, which is considered as a simple model of a biological molecular motor. Furthermore, applications of our result to experimental studies of molecular motors are discussed, and, as an illustration, we predict the value of a new time constant regarding the motion of a KIF1A, which is a kind of single-headed kinesin.
\end{quote}

\section{Introduction} \label{s.intro}
A molecular motor is a mechano-chemical enzyme that transduces the free energy derived from fuel molecules into the mechanical energy of its motion. Many types of molecular motors are known, and they play various roles in living organisms. Here, we overview the variety of known molecular motors and manners in which they function in living cells.

One of the most common molecular motors is an acto-myosin system, which plays the key role in muscle contraction \cite{Hill:1974}.
The acto-myosin system consists of two types of filaments: a myosin (thick) filament and an actin (thin) filament. The myosin filament comprises head and tail parts, and the former plays an important role in the function of the system. The myosin head has enzymatic activity and catalyzes hydrolysis reactions of an adenosine tri-phosphate (ATP) into an adenosine di-phosphate (ADP) and an inorganic phosphate (Pi) as $\mathrm{ATP} + \mathrm{H_2 O} \to \mathrm{ADP} + \mathrm{Pi}$. The difference between the free energies of the reactants and products induces a relative sliding movement of myosin and actin filaments, and is finally transduced into work performed against external load or the viscous resistance of the surrounding medium.
In the case of acto-myosin, this movement leads to a contraction of a muscle fibril, which is a bundle of acto-myosin filaments.
Because muscle fibrils are packed in parallel to form a muscle, the contraction of muscle fibrils finally causes the contraction of a muscle.
In addition to muscle contraction, myosins and actins are also known to be involved in migration of motile cells and also in the formation of a contraction ring in the case of cell division \cite{Cell}.

Kinesin is another typical example of a molecular motor \cite{Vale:1985}. This molecule also hydrolyzes ATP molecules and steps along a protein filament, termed a microtubule. One of the main roles of this molecule is to carry organelles, such as lipid vesicles, over a long distance in a cell. For instance, a kinesin carries an organelle along the axon of a neuron, whose length is sometimes on the order of meters. Because it takes an extremely long time for organelles to be transported by diffusion alone over such a distance, active transporters, such as kinesins, are necessary to allow the tip of an axon communicate with the cell body.

Although the above two species are linear motors, i.e., the motors that slide on a one-dimensional filament, rotating motors have also been found.
$\mathrm{F_o}$-$\mathrm{F}_1$ ATP synthase is a large complex of proteins that is found on an inner membrane of mitochondria \cite{Boyer:1993}.
This motor is known to be composed of two parts: $\mathrm{F_o}$ and $\mathrm{F}_1$. $\mathrm{F_o}$ is embedded in the membrane and it forms a channel, through which protons ($\mathrm{H}^+$) flow according to the difference in their electrochemical potential across the membrane.
This flow of protons is converted into a rotation of $\mathrm{F_o}$ around an axis of the molecule.
This might be understood by analogy with a windmill. $\mathrm{F}_1$ is mechanically connected to $\mathrm{F_o}$, and they share the same rotating axis. The rotation of $\mathrm{F_o}$ induces that of $\mathrm{F}_1$, and the mechanical energy of this rotation is in turn used by $\mathrm{F}_1$ to catalyze the synthesis of ATPs as $\mathrm{ADP} + \mathrm{Pi} \to \mathrm{ATP} + \mathrm{H_2 O}$. In this fashion, the electrochemical potential difference of protons is transduced to the chemical free energy of ATPs.
It is believed that the above scenario can be rewound; it is believed that if the concentration of ATPs is much larger than those of their hydrolysis products, $\mathrm{F_o}$-$\mathrm{F}_1$ is forced to rotate in the opposite direction, and protons are pumped against the gradient of their concentrations. Indeed, counter rotation of $\mathrm{F}_1$ in the presence of surplus ATP has already been observed in experiments \cite{Noji:1997}.

Besides these examples, a wide variety of molecular motors are already known, and many motors are yet to be discovered. However, they share the same property; they catalyze the hydrolysis reaction of ATPs and convert the free energy derived from this reaction into the mechanical energy of their motion. Their dimensions are around ten nanometers, and their movements are quite stochastic.

In the past decade, it has become possible to directly observe the stochastic motion of a single molecular motor due to developments in experimental techniques \cite{Noji:1997, Svoboda:1993, Kitamura:1999}.
In the single-molecule experiments of this kind, the resolutions in space and time are around one nanometer and one millisecond, respectively. These high resolutions enable us to detect every step in a trace of a movement of a motor.
Using such experimental data, kinetic properties of the stochastic movement of molecular motors have been investigated \cite{Svoboda:1994, Astumian:1994, Nishiyama:2002, Carter:2005}.

In contrast to the kinetics, the energetic characteristics of molecular motors remain unclear. For example, there is no systematic method at present to determine the amount of free energy consumed by a single motor at the single-molecule level. Furthermore, we do not know how to determine the amount of the input energy delivered to each mode of the movement of the motor, although this information is quite important in order to elucidate the mechanism for the motion of the motor.
Therefore, it is necessary to establish a systematic method to investigate the energetic properties of molecular motors on the basis of single-molecule experiments.


Before considering how we can achieve this goal, we review the recent theoretical developments in modeling a molecular motor.
In the past ten years, a number of theoretical models for molecular motors have been developed.
In particular, a class of nonequilibrium Langevin systems has been studied intensively, which are termed {\it Brownian motor models} \cite{Julicher:1997, Reimann:2002}. Here, we describe an often referred example of a Brownian motor model, termed {\it a flashing ratchet}, in detail \cite{Astumian:1994, Prost:1994}.
This model describes the dynamics of a colloidal particle in one-dimensional space. This particle is supposed to have several internal states, and transitions from one state to another occur stochastically at certain transition rates. This particle feels a potential energy that depends on the internal state of the particle. 
It has been found in this model that this particle will acquire non-zero mean velocity in general, although the particle is subject to the periodic potentials alone and no macroscopic potential gradient is present.
This is related to the fact that the detailed-balance condition is not satisfied when the inter-state transitions occur, because the transitions make the probability distribution in a steady state different from the canonical (Gibbs-Boltzmann) distribution. When the detailed-balance condition is violated in this manner, the system is not in equilibrium, but in a nonequilibrium steady state.
For a system in a nonequilibrium steady state, a directed motion of the particle is not prohibited by the second law of thermodynamics if there is no spatial reflection symmetry in the system. This effect is a consequence of Curie's principle and often referred to as {\it the ratchet effect} \cite{Prost:1994}. See also the discussion in Sec.~\ref{s.model}.


Energetics of Brownian motor models have also been considered. Sekimoto provided a foundation of energetic considerations of Langevin systems in a general context \cite{Sekimoto:1997}. The formulation developed by Sekimoto has been used to calculate the energetic efficiency of various Brownian motor models \cite{Kamegawa:1998, Takagi:1999}. Parrondo {\it et al.}~also calculated the rate of input energy for several Brownian motor models, including the flashing ratchet \cite{Parrondo:1998}. According to their method, the rate of energy input can be calculated provided that the model parameters such as the profiles of periodic potentials and transition rates are given.


Considering these developments, it seems possible to calculate the rate of energy input of a molecular motor, if we adopt a Brownian motor model as a model of the motor molecule and determine model parameters from experimental data. Unfortunately, however, it is  usually difficult to determine all the model parameters on the basis of experimental data because  the model has too many parameters. Even if we could determine all the model parameters by fitting the calculated quantities from the model to experimental data, validity of the values of these parameters can hardly be verified. As a consequence, the energetic considerations based on the experimentally inferred parameters become unreliable.


In order to remove this difficulty, the authors recently proposed a method by which one can determine the rates of energy input and dissipation of a Brownian motor model without knowing the model details \cite{Harada:2005, Harada:2006}. It has been found that the rate of energy dissipation into the heat bath with which the system is in contact is related in an equality to the quantities such as the mean velocity, the correlation function of velocity fluctuations and the response function of the velocity to an external perturbation to the system.
The important point is that this equality involves experimentally measurable quantities alone. Therefore, we can determine the dissipation rate without referring to the details of the model.
Furthermore, since the rate of energy dissipation is balanced with that of energy input for a system in a steady state, the rate of energy input can also be determined in the same manner.
Hence, if the assumption that a Brownian model is relevant to a molecular motor is correct, with aid of this result, one may be able to determine the rate of energy input for a molecular motor by use of the single-molecule experimental techniques.

This result is also significant from the viewpoint of nonequilibrium statistical mechanics. Our result reveals that the rate of energy dissipation is expressed as an extent of violation of the fluctuation-response relation (FRR). The FRR is a relation between the correlation function and the response function of a macroscopic variable, which is established for systems in equilibrium as {\it the fluctuation-dissipation theorem}. The fluctuation-dissipation theorem is proved on the basis of a condition termed the detailed-balance condition that is satisfied in equilibrium. In contrast, the detailed-balance condition is not satisfied at a nonequilibrium steady state, and this leads to violation of the FRR in a system at a nonequilibrium steady state. According to our result, the extent of violation of the FRR is directly related to the rate of energy dissipation into the heat bath, and it serves as a useful measure to characterize a `distance' from an equilibrium state.

In this paper, we describe this result in detail with regard to the flashing ratchet model as an example. With respect to the FRR, several observable quantities are introduced for this model. Then, it is numerically found that violation of the FRR is related to the energy input and dissipation rates in a simple relation.
Furthermore, we mathematically prove these results by formulating a path-integral representation for the flashing ratchet model.
In addition, we discuss applicability of this result to experimental studies of molecular motors. Several technical points that would be problematic in experiments are addressed. In order to illustrate our method, experimental data for KIF1A, which is a kind of kinesins, are analyzed and a parameter related to the diffusive motion of the motor is predicted.

This paper is organized as follows.
First, the flashing ratchet model is introduced in Sec.~\ref{s.model}. Then, the fluctuation-dissipation theorem and its violation in this model are described and numerically demonstrated in Sec.~\ref{s.FDT}. Then, we state our main claim on the relation between violation of the FRR and the  rate of energy dissipation in Sec.~\ref{s.main} and numerically demonstrate it with the flashing ratchet model.
In Sec.~\ref{s.math}, mathematical proofs of the above results are presented.
In Sec.~\ref{s.diss}, several issues related to experimental applications of our method and concluding remarks are presented.

\section{Model} \label{s.model}

\begin{figure}[htbp]
\begin{center}
\includegraphics{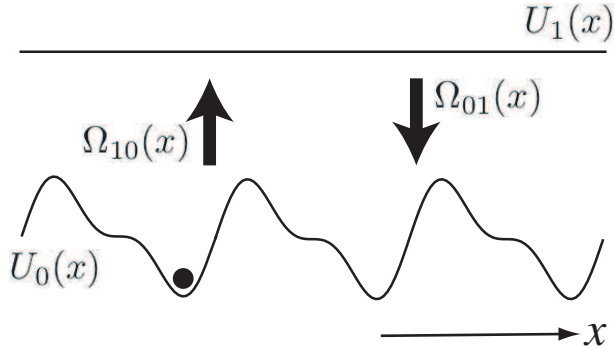}
\caption{Schematic image of the flashing ratchet model. The particle has two internal states, 0 and 1, according to which it feels different periodic potentials, $U_0 (x)$ and $U_1 (x)$, respectively. The internal state of the particle is stochastically switched from 0 to 1 and 1 to 0 at transition rates $\Omega_{10} (x)$ and $\Omega_{01} (x)$, respectively.}
\label{f.model}
\end{center}
\end{figure}

In this Section, we define the flashing ratchet model (See Fig.~\ref{f.model}). Next, several well known results on this model will be briefly reviewed.

Let us consider a Brownian particle suspended in an aqueous solution of temperature $T$.
Motion of this particle is restricted on a one-dimensional track and its position at time $t$ is denoted by $x(t)$.
We suppose this particle is subjected to a force derived from a periodic potential, $U_\sigma (x)$, which is switched from one to another according to the internal state of the particle denoted by $\sigma(t)$. In the following, we assume the internal state, $\sigma(t)$, takes a value either of 0 or 1.
$\sigma(t)$ is supposed to represent a `chemical state' of the particle, and the value of $\sigma(t)$ is stochastically changed from 0 to 1 and vice versa with probabilities per unit time, i.e., the transition rates, $\Omega_{10}(x)$ and $\Omega_{01} (x)$, respectively, which may depend on the current position of the particle.

The equation of motion for this particle is a Langevin equation with the above ingredients as
\begin{equation}
m \dert{x(t)}{t} = - \gamma \der{x(t)}{t} - U'_{\sigma(t)} (x(t)) + \xi(t) + \ve \fp(t),
\label{e.udmodel}
\end{equation}
where $m$ is the mass of the particle and the friction coefficient $\gamma$ represents the viscosity effect from the environment.
$\xi(t)$ is Gaussian white noise with mean 0 and variance $2 \gamma \kb T$, which represents the thermal noise from the environment.
As stated above, $U_\sigma(x)$ is an $\ell$-periodic potential for $\sigma = 0, 1$, and $\sigma(t)$ is a Possion process on $\{ 0, 1 \}$ with transition rates $\Omega_{10} (x(t))$ and $\Omega_{01} (x(t))$ for transitions from 0 to 1 and 1 to 0, respectively.
The prime denotes differentiation with respect to $x$.
The last term in Eq.~(\ref{e.udmodel}) is added here to represent a perturbation force, which is applied from outside. $\ve$ is a dimensionless parameter that is supposed to be much smaller than unity, which assures us of smallness of the external force.

As usual, since the inertial effect is negligible for a small particle in a viscous solvent, we consider the overdamped version of Eq.~(\ref{e.udmodel}) in the following:
\begin{equation}
\gamma \dot x(t) = - U'_{\sigma(t)} (x(t)) + \xi(t) + \ve \fp(t),
\label{e.model}
\end{equation}
where the dot denotes the time derivative: $\dot x(t) \equiv \d x(t) / \d t$.

For the model given in Eq.~(\ref{e.model}), by use of standard techniques \cite{Risken:1996}, it can be proved that the probability distribution $P_\sigma (x, t)$, which represents the probability to find the system in the state $\sigma$ and the position $x$ at time $t$, obeys the following Fokker-Planck equation \cite{Astumian:1994, Prost:1994}.
\begin{equation}
\pder{}{t} \vtr{P_0 (x, t)}{P_1 (x, t)} = \mtx{\LFP_0 (t) - \Omega_{01}(x) }{\Omega_{10}(x)}{\Omega_{01}(x)}{\LFP_1(t) - \Omega_{10}(x)} \vtr{P_0 (x, t)}{P_1 (x, t)},
\label{e.FP}
\end{equation}
where 
\begin{equation}
\LFP_\sigma (t) \equiv \frac{1}{\gamma} \pder{}{x} \left[ U'_\sigma (x) - \ve \fp(t) + \kb T \pder{}{x} \right]
\label{e.FPop}
\end{equation}
is the Fokker-Planck operator for each state.

The steady-state distribution in the absence of perturbation, $\Ps_\sigma(x)$, is obtained from Eq.~(\ref{e.FP}) as a stationary solution with $\ve = 0$.
It has been found that the canonical distribution 
\begin{equation}
\Pc_\sigma (x) = \frac{1}{Z} \exp(- U_\sigma(x)/\kb T),
\label{e.canonical}
\end{equation}
where $Z \equiv \int_0^\ell \sum_{\sigma = 0}^1 \exp(- U_\sigma(x)/\kb T) \d x$, provides the steady-state distribution if and only if the transition rates satisfy the condition
\begin{equation}
\frac{\Omega_{10}(x)}{\Omega_{01}(x)} = \e^{-\left[ U_1(x) - U_0 (x) \right]/ \kb T}.
\label{e.db}
\end{equation}
In this case, as explained in Sec.~\ref{s.ldb}, the detailed-balance condition is satisfied, and the system is in equilibrium.
In general, the steady-state distribution differs from the canonical distribution when Eq.~(\ref{e.db}) is not satisfied, and the system is at a nonequilibrium steady state.
In this case, the detailed-balance condition is violated, which results in the non-vanishing velocity.

Analysis of Eq.~(\ref{e.FP}) reveals that the mean velocity with $\ve = 0$ is expressed as \cite{Prost:1994}
\begin{equation}
\bar v \equiv - \frac{1}{\gamma} \int_0^\ell \left[ \Ps_0 (x) U'_0 (x) + \Ps_1 (x) U'_1 (x) \right] \d x.
\label{e.vs}
\end{equation}
From this expression, it is easily confirmed that the steady velocity $\bar v$ vanishes in the case of equilibrium, where $\Ps_\sigma (x)$ equals the canonical distribution $\Pc_\sigma(x)$. In nonequilibrium steady states, since $\Ps_\sigma (x)$ is different from the canonical distribution, $\bar v$ differs from zero in general.


Next, let us consider energetics of this model.
In the following discussion, we set $\ve = 0$ for a while.
As long as the particle is moving on a single potential, i.e., it stays in a single state, no net energy supply from outside is required because the particle is subject to the potential (conservative) force alone. Injection of energy is necessary when the particle undergoes a transition from one state $\sigma$ to another state $\sigma'$. In this occasion, the difference in potential energies, $U_{\sigma'} (x(t)) - U_{\sigma}(x(t))$, must be externally supplied.
Thus, by representing the timings at which transitions occur during time interval from 0 to $t$ as $\{t_1, t_2 \cdots, t_n \}$, the external energy input during this interval is represented as
\begin{equation}
E(t) = \sum_{j=1}^n \left[ U_{\sigma(t_j + 0)} (x(t_j)) - U_{\sigma(t_j - 0)} (x(t_j)) \right],
\label{e.input}
\end{equation}
which is a stochastic variable.
The average energy input per unit time is calculated as
\begin{eqnarray}
J &\equiv& \lim_{t \to \infty} \frac{E(t)}{t} \nonumber \\
&=& \int_0^\ell \left[ U_1(x) - U_0 (x) \right] \left[ \Omega_{10}(x) \Ps_0(x) - \Omega_{01}(x) \Ps_1(x) \right] \d x.
\label{e.incalc}
\end{eqnarray}
The second line follows from the consideration that the probability per unit time at which a transition from $\sigma$ to $\sigma'$ at $x$ occurs is equal to $\Omega_{\sigma' \sigma}(x) \Ps_\sigma (x)$ at a steady state.
As mentioned in Sec.~\ref{s.intro}, calculation of input and dissipated energy, $J$, using Eq.~(\ref{e.incalc}) requires the detailed knowledge on the potential profiles and the transition rates.

The above definition of input energy is compatible with Sekimoto's argument on energetics of Langevin systems.
Sekimoto suggested that dissipated energy into the environment during time interval from 0 to $t$  can be expressed as
\begin{equation}
Q(t) = \int_0^t \left[ \gamma \dot x(s) - \xi(s) \right] \circ \d x(s),
\label{e.heat}
\end{equation}
where the symbol $\circ$ means that the above Stieltjes integral should be interpreted according to Stratonovich's prescription \cite{Gardiner:2004}.
$Q(t)$ is also a stochastic variable.
With aid of Eqs.~(\ref{e.model}) and (\ref{e.input}), we obtain the following relation
\begin{equation}
E(t) = Q(t) + U_{\sigma(t)} (x(t)) - U_{\sigma(t)} (x(0)),
\label{e.balance}
\end{equation}
which expresses the conservation law of energy.
Since $\left[ U_{\sigma(t)} (x(t)) - U_{\sigma(0)} (x(0)) \right]/t \to 0$ holds as $t \to \infty$ at a steady state, the average rate of energy input is equal to that of energy dissipation as $\lim_{t \to \infty} E(t)/t = \lim_{t \to \infty} Q(t)/t = J$.

\section{The Fluctuation-Response Relation} \label{s.FDT}

In this Section, we introduce the fluctuation-response relation (FRR) with regard to the flashing ratchet model.
First, after defining several quantities, the fluctuation-dissipation theorem is stated.
We then demonstrate the FRR numerically for the flashing ratchet model in equilibrium.
Next, we numerically show that the FRR is violated in a nonequilibrium steady state of the flashing ratchet model.

\subsection{Equilibrium case: The Fluctuation-Dissipation Theorem}

First, let us define several quantities that are involved in the statement of the theorem.
One of the important quantities is the velocity correlation function defined as follows.
\begin{equation}
C(t) \equiv \bra \left[ \dot x(t) - \bar v \right] \left[ \dot x(0) - \bar v \right] \ket_0,
\label{e.corr}
\end{equation}
where $\bra \cdots \ket_0$ denotes the ensemble average in the absence of a perturbation ($\ve = 0$).
According to Wiener-Khinchin theorem, the Fourier transform of $C(t)$ defined as
\begin{equation}
\tilde C(\omega) = \int_{-\infty}^\infty C(t) \e^{i \omega t} \d t,
\label{e.psd}
\end{equation}
is equal to the power spectrum density of the velocity fluctuations, $\dot x(t) - \bar v$.
Because a system in a steady state possesses the time translational invariance, we find a symmetry property as $C(t) = C(-t)$. The imaginary part of $\tilde C(\omega)$ vanishes because of this symmetry.

Next, we define the response function that characterizes the linear response of the velocity to a perturbation force as
\begin{equation}
\lim_{\ve \to 0} \frac{\bra \dot x(t) \ket_\ve - \bar v}{\ve} = \int_{-\infty}^t R(t-s) \fp(s) \d s,
\label{e.res}
\end{equation}
where $\bra \cdots \ket_\ve$ denotes the ensemble average taken in the presence of the perturbation. 
By adopting the perturbation $\fp (s)$ as an impulse at time 0, i.e., $\fp (s) = \delta(s)$, it is found that the response function $R(t)$ describes the response of the mean velocity to this impulse as
\begin{equation}
R(t) = \lim_{\ve \to 0} \frac{\bra \dot x(t) \ket_\ve - \bar v}{\ve} \quad \mathrm{with}\quad \fp(t) = \delta (t).
\label{e.res2}
\end{equation}
Because of the causality, $R(t)$ for negative $t$ is always zero.

Due to the convolution theorem, the Fourier transform of the response function, termed {\it susceptibility},
\begin{equation}
\tilde R(\omega) = \int_0^\infty R(t) \e^{i \omega t} \d t
\label{e.sus}
\end{equation}
satisfies the following relation
\begin{equation}
\tilde R(\omega)  = \frac{1}{\tilde \fp(\omega)} \lim_{\ve \to 0} \int_0^\infty \frac{\bra \dot x(t) \ket_\ve - \bar v}{\ve} \e^{i \omega t} \d t,
\end{equation}
where $\tilde \fp(\omega)$ is the Fourier transform of $\fp(t)$.
This relation enables us to determine the susceptibility experimentally by applying a sinusoidal perturbation: $\fp(t) = \sin \omega t$.

Now, we can state the fluctuation-dissipation theorem.
Here, we suppose that Eq.~(\ref{e.db}) is satisfied and thus the system without perturbation ($\ve = 0$) is in equilibrium.
In this case, as mentioned above, the mean velocity of the particle vanishes: $\bar v = 0$.
Furthermore, the correlation function and the response function satisfy the following relation
\begin{equation}
C(t) = \kb T R(t) \quad \mathrm{for}\quad t > 0,
\label{e.tFDT}
\end{equation}
or equivalently
\begin{equation}
\tilde C(\omega) = 2 \kb T \tilde R^{+} (\omega),
\label{e.fFDT}
\end{equation}
where $\tilde R^{+}(\omega)$ represents the real part of $\tilde R(\omega)$.
Proof of these relations is provided in Sec.~\ref{s.math}.
In what follows, we term Eq.~(\ref{e.fFDT}) the fluctuation-response relation (FRR).

It should be noted that the FRR exactly holds independent of the model datails such as the profiles of periodic potentials and the transition rates.
As explained in Sec.~\ref{s.FDTproof}, the FRR is a direct consequence of the detailed-balance property of a system in equilibrium.
Due to this simple relation, the measurements of the correlation function and the response function give the identical information on the system if it is in equilibrium. 

\begin{figure}[htbp]
\begin{center}
\includegraphics{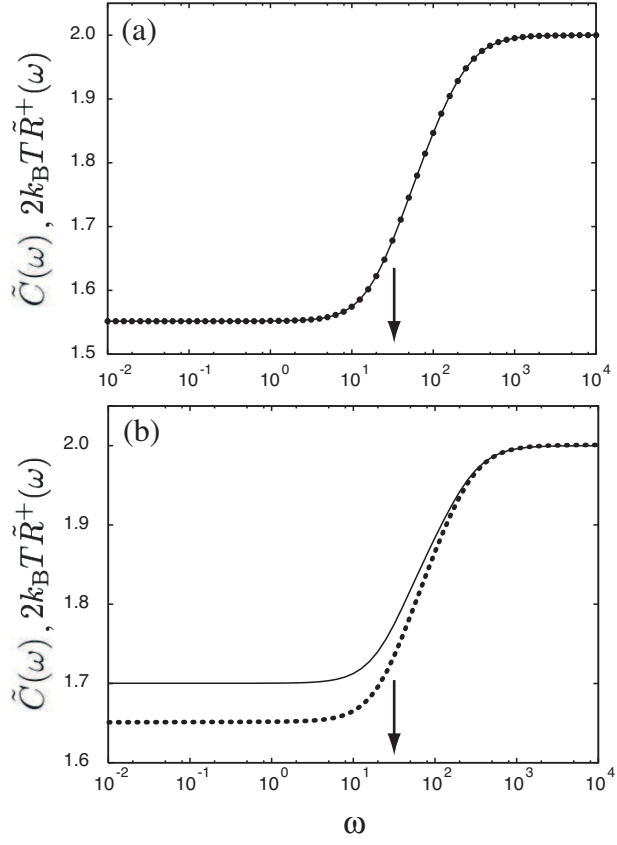}
\caption{$\tilde C(\omega)$ and $2 \kb T\tilde R^{+} (\omega)$ as functions of  $\omega$.
(a) Equilibrium case ($\Omega_{10} (x) = \Omega_{01} (x) = 0$). (b) Nonequilibrium case ($\Omega_{10} (x) = \Omega_{01} (x) = 10$).
The solid and dotted curves represent $\tilde C(\omega)$ and $2 \kb T\tilde R^{+} (\omega)$, respectively. The arrows donote the decay rate $\omega^*$.
All the quantities are dimensionless under the normalization as $\ell = \gamma = \kb T = 1$.}
\label{f.FDT}
\end{center}
\end{figure}

Now, we numerically demonstrate the FRR for the flashing ratchet model.
For the numerical calculation, we adopted the model parameters as $U_0 = \sin (2 \pi x/\ell) + 1/2 \sin (4 \pi x/\ell)$ and $U_1 = const.$, $\Omega_{10} (x) = \Omega_{01} (x) = 0$ \footnote{This is a special case of Eq.~(\ref{e.db}). These transition rates are obtained by choosing $\Omega_{10} (x) = \Omega \e^{\beta U_0 (x)}$ and $\Omega_{01} (x) = \Omega \e^{\beta U_1(x)}$ and letting $\Omega \to 0$.}. All the quantities are made dimensionless by normalizing $\ell  = \gamma = \kb T = 1$.
On the basis of model Eqs.~(\ref{e.model}) and (\ref{e.FP}), we calculated the power spectrum density $\tilde C(\omega)$ and the susceptibility $\tilde R(\omega)$ by using the Fourier-modes expansion \cite{Harada:2005a, Risken:1996}.
In Fig.~\ref{f.FDT} (a), $\tilde C(\omega)$ and $2 \kb T \tilde R^{+} (\omega)$ are plotted as functions of the frequency $\omega$. As clearly seen, the FRR holds irrespective of the frequency, implying Eq.~(\ref{e.fFDT}). Since Eq.~(\ref{e.tFDT}) can be obtained by inverse Fourier transform of Eq.~(\ref{e.fFDT}), Eq.~(\ref{e.tFDT}) has also been verified.

We can obtain several important quantities from this plot.
First, the diffusion coefficient, defined as
\begin{equation}
D \equiv \lim_{t \to \infty} \frac{ \left[ x(t) - x(0) - \bar v t \right]^2 }{2t},
\label{e.diff_c}
\end{equation}
satisfies $D = \tilde C(0)/2$. Therefore, we can read the value of $D$ from the low-frequency plateau of $\tilde C(\omega)$.
Next, the mobility, which characterizes the linear response of the velocity to a constant external force $\fp(t) = \fp$ as
\begin{equation}
\mu \equiv \lim_{\ve \to 0} \frac{\bra \dot x \ket_\ve - \bar v}{\ve \fp},
\end{equation}
is similarly determined through the relation $\mu = \tilde R^{+}(0)$.

The bare friction coefficient $\gamma$, which appears in the model equation (\ref{e.model}) can be determined from the high-frequency plateau of $\tilde R^{+}(\omega)$, because it satisfies $\gamma = \lim_{\omega \to \infty} 1/\tilde R^{+}(\omega)$.

One may notice that the two plateaus are separated at an intermediate frequency, $\omega^*$ (indicated by an arrow in Figs.~\ref{f.FDT} (a)). This frequency corresponds to the decay rate of the probability distribution inside a potential well.
Roughly speaking, this rate is the inverse of the time needed for a particle to diffuse around the bottom of the potential well.
More precisely, this frequency coincides with the smallest absolute value of non-vanishing eigenvalues of the Fokker-Planck operator given in Eq.~(\ref{e.FPop}).

Note that these correspondences are not restricted to a system in equilibrium. The above-mentioned quantities can be obtained for nonequilibrium systems in the same manner.

\subsection{Nonequilibrium case: Violation of Fluctuation-Response Relation} \label{ss.viol}

Next, let us examine what happens in a nonequilibrium case.
We conducted the same calculation as above but with different parameter values: $\Omega_{10} (x) = \Omega_{01} (x) = 10~(\gamma \ell^2/T)$.
In this case, the condition given in Eq.~(\ref{e.db}) is not satisfied and the canonical distribution is not the steady-state distribution.
Then, from Eq.~(\ref{e.vs}), the mean velocity does  not vanish in general as mentioned earlier.
Actually, if we employ the same potential profiles as above, $U_0 = \sin (2 \pi x/\ell) + 1/2 \sin (4 \pi x/\ell)$ and $U_1 = const.$, we numerically obtain $\bar v = 0.054~T/(\gamma \ell)$.
Furthermore, we find that the FRR [Eq.~(\ref{e.fFDT})] is violated in this case as depicted in Fig.~\ref{f.FDT} (b). In this figure, violation of the FRR is evident in particular for small frequencies. For the frequencies much larger than $\omega^*$, the FRR is retained even in a nonequilibrium steady state.
This can be qualitatively explained as follows; For a timescale much shorter than $1/\omega^*$, the particle does not {\it know} whether the potential is being switched because it does not travel a long distance within this short time range and the particle effectively feels no potential. Therefore, the statistical properties of the motion of the particle in the presence of the switching potential become same as those of a free diffusion of a particle for these short time scales, and the FRR holds. However, for a longer timescale, the particle starts to feel the potential as well as its switching. Then, the statistical property deviates from that of equilibrium system, and the FRR is violated.
In the next Section, we consider the physical significance of violation of the FRR.

\section{Main Claim} \label{s.main}
In this Section, we state our main claim.
In the previous Section, it is found that the FRR is violated in a nonequilibrium steady state.
Such a nonequilibrium steady state is maintained by the external energy input and energy dissipation into the environment.
Then, by comparing the amounts of energy flows, $J$ [see Eq.~(\ref{e.incalc})], with the degrees of the FRR violation in nonequilibrium steady states, we are led to a simple relation between these quantities. This relation is expressed by the following equality
\begin{equation}
J = \gamma \left\{ \bar v^2 + \int_{-\infty}^\infty \left[ \tilde C(\omega) - 2 \kb T \tilde R^{+}(\omega) \right] \frac{\d \omega}{2\pi} \right\}.
\label{e.eq}
\end{equation}
This equality is proved in Sec.~\ref{s.math}.

Interestingly, the right-hand side of Eq.~(\ref{e.eq}) expresses the extent of the FRR violation. The first term in the braces, the square mean velocity, deviates from zero when the detailed-balance condition is violated in nonequilibrium states. The second term in the braces represents the extent of violation of Eq.~(\ref{e.fFDT}) over the frequency domain. This quantity is precisely equal to the area of the region surrounded by two curves $\tilde C(\omega)$ and $2 \kb T \tilde R^{+} (\omega)$ in a plot of them as functions of $\omega$ [see Fig.~\ref{f.FDT} (b)].
Notice that the square-mean velocity can be interpreted as the zero-frequency component of the velocity correlation, $\bra \dot x(t) \dot x(0) \ket_0$.
Therefore, the right-hand side of Eq.~(\ref{e.eq}) is interpreted as the sum of dissipated energy per unit time through each frequency of velocity fluctuations.
In equilibrium, since the mean velocity vanishes and Eq.~(\ref{e.fFDT}) holds, the right-hand side of Eq.~(\ref{e.eq}) vanishes.
This is consistent with the fact that the rate of energy input and dissipation, $J$, vanishes in equilibrium.

One of the outstanding features of Eq.~(\ref{e.eq}) is that it is independent of the system details such as the profiles of the potentials and the transition rates.
Indeed, this equality holds regardless of these model parameters. It is also independent of the number of the internal states.
Furthermore, this equality can be proved for a wide variety of Langevin models other than the flashing ratchet model. It can be extended to time-dependent systems and systems which include multiple heat reservoirs. Therefore, this type of relation between the energy dissipation rate and the extent of the FRR violation is universal for a class of Langevin systems. See Ref.~\cite{Harada:2006} for further details.

\begin{figure}[htbp]
\begin{center}
\includegraphics{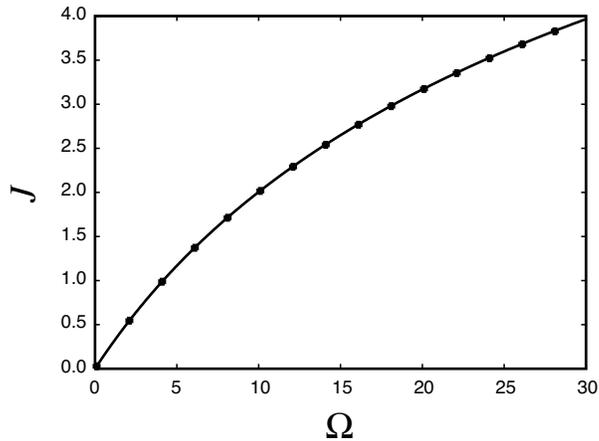}
\caption{Numerical verification of Eq.~(\ref{e.eq}) for the flashing ratchet model.
The solid curve represents the values of $J$ calculated from Eq.~(\ref{e.incalc}). The filled squares are values of the right-hand side of Eq.~(\ref{e.eq}). The horizontal axis represents the values of the transition rates $\Omega$.
All the quantities are dimensionless under the normalization as $\ell = \gamma = \kb T = 1$.}
\label{f.main}
\end{center}
\end{figure}

We now numerically demonstrate Eq.~(\ref{e.eq}) for the flashing ratchet model.
In Fig.~\ref{f.main}, the energy input rate $J$ calculated as in Eq.~(\ref{e.incalc}) and the right-hand side of Eq.~(\ref{e.eq}) are plotted independently as functions of the transition rate $\Omega$ [The transition rates were taken uniform as $\Omega = \Omega_{10} (x) = \Omega_{01} (x)$].
As clear seen in this figure, Eq.~(\ref{e.eq}) holds for all the values of the transition rate $\Omega$.

\section{Mathematical Proofs} \label{s.math}
In this Section, mathematical proofs of the result described in the former Sections are presented.

\subsection{Path-integral representation of the flashing ratchet model} \label{s.path}

As a preparation, in this Section, we develop a path-integral representation of the flashing ratchet model expressed as
\begin{eqnarray}
\pder{}{t} \vtr{P_0 (x, t)}{P_1 (x, t)} = \LFP(t) \vtr{P_0 (x, t)}{P_1 (x, t)},
\label{e.FP2}
\end{eqnarray}
where
\begin{equation}
\LFP(t) \equiv \frac{1}{\gamma} \pder{}{x} \mtx{U'_0 (x)}{0}{0}{U'_1 (x)} - \frac{\ve}{\gamma} \fp(t) \pder{}{x} + \frac{T}{\gamma} \pdert{}{x} + \mtx{- \Omega_{10}(x)}{\Omega_{01} (x)}{\Omega_{10}(x)}{- \Omega_{01}(x)},
\label{e.FPop2}
\end{equation}
In this Section, the Boltzmann constant is normalized to unity.

\subsubsection{Definitions and Notations}

Let us consider a standard description employed in quantum mechanics.
We define the position and momentum operators as
\begin{eqnarray}
\mathcal{X} &\equiv& x \\
\mathcal{P} &\equiv& -i \pder{}{x},
\end{eqnarray}
and denote eigenvectors of these operators as 
\begin{eqnarray}
\mathcal{X} \lket{x} &=& x \lket{x} \\
\mathcal{P} \lket{p} &=& p \lket{p}.
\end{eqnarray}
These eigenvectors are orthonormalized 
\begin{eqnarray}
\braket{x}{x'} &=& \delta(x - x') \\
\braket{p}{p'} &=& \delta(p - p'), 
\end{eqnarray}
and regarded as complete sets in the state space $\mathfrak{H}$
in the sense that 
\begin{eqnarray}
\int_{-\infty}^\infty \d x \lket{x} \bral{x} = \int_{-\infty}^\infty \d p \lket{p} \bral{p} = \mathcal{I}, 
\end{eqnarray}
where $\mathcal{I}$ is an identity operator. 
We can also confirm the following relations
\begin{eqnarray}
\braket{x}{p} &=& \frac{1}{\sqrt{2\pi}} \e^{i x p}, \\
\braket{p}{x} &=& \frac{1}{\sqrt{2\pi}} \e^{-i x p}.
\end{eqnarray}
Next, we consider a vector space, $\mathbb{C}^2$, and the following complete set in $\mathbb{C}^2$
\begin{eqnarray}
\lket{0} &\equiv& \vtr{1}{0} \\
\lket{1} &\equiv& \vtr{0}{1}
\end{eqnarray}
We then consider the product space $\mathbb{C}^2 \times \mathfrak{H}$, whose element can be expressed as $\lket{\sigma, x} \equiv \lket{\sigma} \lket{x}$ with $\sigma \in \{0, 1\}$. $\lket{\sigma, x}$ spans a complete set in $\mathbb{C}^2 \times \mathfrak{H}$ as
\begin{equation}
\int_{-\infty}^\infty \d x \left( \lket{0, x} \bral{0, x} + \lket{1, x} \bral{1, x} \right) = \mathcal{I}.
\label{e.comp}
\end{equation}
Using the above notations, $\LFP(t)$ defined in Eq.~(\ref{e.FPop2}) can be expressed as an operator acting on $\mathbb{C}^2 \times \mathfrak{H}$ as follows
\begin{eqnarray}
\LFP(t) &=& \frac{i}{\gamma} \mathcal{P} \left[ U'_0  (\mathcal{X}) \lket{0} \bral{0} + U'_1 (\mathcal{X}) \lket{1} \bral{1} \right] - \frac{i \ve}{\gamma} \fp(t) \mathcal{P} - \frac{T}{\gamma} \mathcal{P}^2  \nonumber \\
& & - \Omega_{10} (\mathcal{X}) \left( \lket{0}\bral{0} - \lket{1}\bral{0} \right) + \Omega_{01} (\mathcal{X}) \left( \lket{0} \bral{1} - \lket{1} \bral{1} \right)
\label{e.op}
\end{eqnarray}

\subsubsection{Representation}
Next, we obtain a representation of $\LFP(t)$ in terms of the position.
We first use the following identity
\begin{equation}
\LFP(t) = \int \!\!\! \int \!\!\! \int \!\!\! \int \lket{x_1} \bral{p_1} \LFP(t) \lket{p_2} \bral{x_2} \frac{1}{2\pi} \e^{i(x_1 p_1 - x_2 p_2)} \d x_1 \d x_2 \d p_1 \d p_2.
\end{equation}
The following transformation of variables
\begin{eqnarray}
\left\{ \begin{array}{rcl}
x &=& \frac{1}{2} (x_1 + x_2) \\
y &=& x_1 - x_2 \\
p &=& \frac{1}{2} (p_1 + p_2) \\
q &=& p_1 - p_2
\end{array} \right.
\end{eqnarray}
yields
\begin{equation}
\LFP(t) = \int \!\!\! \int \!\!\! \int \lket{x + \frac{1}{2} y} \bral{x - \frac{1}{2} y} L(x, p, t) \frac{1}{2\pi} \e^{i y p} \d x \d y \d p,
\label{e.op2}
\end{equation}
where
\begin{equation}
L(x, p, t) \equiv \int_{-\infty}^\infty \bral{p + \frac{1}{2} q} \LFP(t) \lket{p - \frac{1}{2} q} \e^{i x q} \d q.
\label{e.L}
\end{equation}
From Eq.~(\ref{e.op2}), we obtain
\begin{equation}
\bral{x} \LFP(t) \lket{x'} = \int_{-\infty}^\infty \frac{\d p}{2\pi} \e^{i p(x - x')} L\left(\frac{x+x'}{2}, p , t\right).
\label{e.repop}
\end{equation}
By substituting Eq.~(\ref{e.op}) into Eq.~(\ref{e.L}), it is calculated
\begin{eqnarray}
L(x, p, t) &=& \frac{1}{\gamma} \left\{ \left[ p U'_0(x) - \frac{i}{2} U''_0 (x) \right] \lket{0}\bral{0} + \left[ p U'_1(x) - \frac{i}{2} U''_1 (x) \right] \lket{1}\bral{1} \right\} - \frac{i \ve}{\gamma} \fp(t) p \nonumber \\
& & - \frac{T}{\gamma}p^2 - \Omega_{10} (x) \left( \lket{0}\bral{0} - \lket{1}\bral{0} \right) + \Omega_{01} (x) \left( \lket{0} \bral{1} - \lket{1} \bral{1} \right)
\label{e.L2}
\end{eqnarray}

\subsubsection{Path-integral formulation}
Next, we integrate Eq.~(\ref{e.FP2}) from $t = t_j$ over a small interval $\Delta t$ as
\begin{eqnarray}
\vtr{P_0(x, t_{j+1})}{P_1 (x, t_{j+1})} &=& \left[ \mathcal{I} + \Delta t \LFP (t_j) \right] \vtr{P_0(x, t_j)}{P_1 (x, t_j)} + O(\Delta t^2) \nonumber \\
&\equiv& \mathcal{U}(t_{j+1}, t_j) \vtr{P_0(x, t_j)}{P_1 (x, t_j)} + O(\Delta t^2),
\label{e.int}
\end{eqnarray}
where $t_{j+1} = t_j + \Delta t$.
Then, we insert Eq.~(\ref{e.comp}) with $x = x_{j+1}$ and $x = x_j$ into the left and the right of $\mathcal{U}(t_{j+1}, t_j)$, respectively.
Further integration and successive insertions of Eq.~(\ref{e.comp}) lead to the following expression of the probability distribution
\begin{equation}
P_{\sigma_N} (x_N, t_N) = \prod_{j = 0}^{N-1} \int_{-\infty}^\infty \d x_j \sum_{\sigma_j \in \{ 0, 1\}} \bral{\sigma_{j+1}, x_{j+1}} \mathcal{U}(t_{j+1}, t_j) \lket{\sigma_j, x_j} P_{\sigma_0} (x_0, t_0),
\label{e.prob}
\end{equation}
where terms of $O(\Delta t^2)$ have been omitted.
From Eq.~(\ref{e.prob}), we obtain the transition probability of a path 
\begin{equation}
[ \sigma, x ] \equiv \{ (\sigma_0, x_0), (\sigma_1, x_1), ..., (\sigma_N, x_N) \}
\end{equation}
which starts $(\sigma_0, x_0)$ at $t = t_0$ as
\begin{equation}
P_\ve ((\sigma_0, x_0) \to [\sigma, x]) = \prod_{j = 0}^{N-1} \bral{\sigma_{j+1}, x_{j+1}} \mathcal{U}(t_{j+1}, t_j) \lket{\sigma_j, x_j}.
\label{e.trans}
\end{equation}

By use of Eqs.~(\ref{e.repop}) and (\ref{e.L2}), we can calculate each factor entering Eq.~(\ref{e.trans}) as
\begin{eqnarray}
\lefteqn{\bral{0, x_{j+1}} \mathcal{U}(t_{j+1}, t_j) \lket{0, x_j} =} \nonumber \\
& &  \sqrt{\frac{\gamma}{4 \pi T \Delta t}} \exp \left\{ -\frac{\Delta t}{4 \gamma T} \left[ \gamma \frac{x_{j+1} - x_j}{\Delta t} + U'_0 (\bar x_j) - \ve \bfp{j} \right]^2 + \frac{\Delta t}{2\gamma} U''_0 (\bar x_j) - \Delta t \Omega_{10} (\bar x_j) \right\},
\end{eqnarray}
where $\bar x_j \equiv (x_j + x_{j+1})/2$ and $\bfp j \equiv [ \fp (t_j) + \fp(t_{j+1}) ]/2$.
Similar calculations yield
\begin{eqnarray}
\lefteqn{\bral{1, x_{j+1}} \mathcal{U}(t_{j+1}, t_j) \lket{1, x_j} =} \nonumber \\
& &  \sqrt{\frac{\gamma}{4 \pi T \Delta t}} \exp \left\{ -\frac{\Delta t}{4 \gamma T} \left[ \gamma \frac{x_{j+1} - x_j}{\Delta t} + U'_1 (\bar x_j) - \ve \bfp{j} \right]^2 + \frac{\Delta t}{2\gamma} U''_1 (\bar x_j) - \Delta t \Omega_{01} (\bar x_j) \right\},
\end{eqnarray}
and
\begin{eqnarray}
\bral{1, x_{j+1}} \mathcal{U} (t_{j+1}, t_j) \lket{0, x_j} &=& \Delta t \Omega_{10} (\bar x_j) \delta(x_{j+1} - x_j), \\
\bral{0, x_{j+1}} \mathcal{U} (t_{j+1}, t_j) \lket{1, x_j} &=& \Delta t \Omega_{01} (\bar x_j) \delta(x_{j+1} - x_j).
\end{eqnarray}

Using the above expressions, we can explicitly write down the transition probability.
More concretely, let us suppose that the state $\sigma$ is switched at time intervals with $j = j_1, j_2, ..., j_n$ as $\lket{\sigma_{j_m +1}} =\left( {{0~1} \atop {1~0}} \right) \lket{\sigma_{j_m}}$ for $m \in \{1, 2, ..., n \}$. Then, $\sigma_j$ for $j_{m} < j \le j_{m+1}$ ($ m \in \{ 0, 1, .., n\}$ and $j_0 \equiv -1$, $j_{n+1} \equiv N$) has the same value represented by $\Sigma_m$. 
The transition probability of this path is represented as
\begin{eqnarray}
\lefteqn{P_\ve ((\sigma_0, x_0) \to [\sigma, x]) = } \nonumber \\
& & \prod_{j = 0}^{j_1 -1} \sqrt{\frac{\gamma}{4 \pi T \Delta t}} \e^{ -\frac{\Delta t}{4 \gamma T} \left[ \gamma \frac{x_{j+1} - x_j}{\Delta t} + U'_{\Sigma_0} (\bar x_j) - \ve \bfp{j} \right]^2 + \frac{\Delta t}{2\gamma} U''_{\Sigma_0} (\bar x_j) - \Delta t \Omega_{\check \Sigma_0 \Sigma_0} (\bar x_j)} \nonumber \\
& & \times \prod_{m = 1}^n \Delta t \Omega_{\Sigma_m \Sigma_{m-1}} (\bar x_{j_m -1}) \delta( x_{j_m} - x_{j_m - 1})  \nonumber \\
& & \times \prod_{j = j_m + 1}^{j_{m+1}-1} \sqrt{\frac{\gamma}{4 \pi T \Delta t}} \e^{ -\frac{\Delta t}{4 \gamma T} \left[ \gamma \frac{x_{j+1} - x_j}{\Delta t} + U'_{\Sigma_m} (\bar x_j) - \ve \bfp{j} \right]^2 + \frac{\Delta t}{2\gamma} U''_{\Sigma_m} (\bar x_j) - \Delta t \Omega_{\check \Sigma_m \Sigma_m} (\bar x_j)},
\label{e.path}
\end{eqnarray}
where $\check \Sigma_m \equiv - (\Sigma_m - 1)$.

\subsection{Proof of the fluctuation-dissipation theorem} \label{s.FDTproof}
In this Section, we prove the fluctuation-dissipation theorem for the flashing ratchet model with aid of the results obtained in Sec.~\ref{s.path}.

\subsubsection{Local Detailed-Balance Condition} \label{s.ldb}
First, we consider the time-reversed path of a path $[\sigma, x]$, denoted by 
\begin{equation}
\hat{[\sigma, x]} = \{(\sigma_N, x_N), (\sigma_{N-1}, x_{N-1}), ..., (\sigma_0, x_0) \}.
\end{equation}
Let $\hat \fp(t)$ represent the time-reversed counterpart of $\fp(t)$ as $\hfp{j} \equiv \bfp{N - j - 1}$. We then define the transition probability of a path $[\sigma, x]$ subject to $\hfp{j}$, which is represented by $\hat P_\ve ((\sigma_0, x_0) \to [\sigma, x])$, by replacing $\bfp{j}$ in Eq.~(\ref{e.path}) with $\hfp{j}$.
With these preparations, we find 
\begin{eqnarray}
\frac{P_\ve ((\sigma_0, x_0) \to [\sigma, x])}{\hat P_\ve ((\sigma_N, x_N) \to \hat{[\sigma, x]})} &=& \e^{\beta U_{\sigma_0} (x_0) -\beta U_{\sigma_N} (x_N) + \beta \ve \sum_{j=0}^{N-1} \bfp{j} (x_{j+1} - x_j)} \nonumber \\
& & \times \prod_{m=1}^n \e^{-\beta U_{\Sigma_{m-1}} (\bar x_{j_m})} \frac{\Omega_{\Sigma_m \Sigma_{m-1}} (\bar x_{j_m})}{\Omega_{\Sigma_{m-1} \Sigma_m} (\bar x_{j_m})} \e^{\beta U_{\Sigma_m} (\bar x_{j_m})},
\label{e.ldb}
\end{eqnarray}
where $\beta \equiv T^{-1}$. If the condition in Eq.~(\ref{e.db}) is satisfied, Eq.~(\ref{e.ldb}) is reduced to
\begin{equation}
\frac{P_\ve ((\sigma_0, x_0) \to [\sigma, x])}{\hat P_\ve ((\sigma_N, x_N) \to \hat{[\sigma, x]})} = \e^{\beta U_{\sigma_0} (x_0) -\beta U_{\sigma_N} (x_N) + \beta \ve \sum_{j=0}^{N-1} \bfp{j} (x_{j+1} - x_j)}.
\label{e.ldb0}
\end{equation}
This expression is a key property used in the proof of the fluctuation-dissipation theorem, and is termed {\it the local detailed-balance condition} \cite{Crooks:2000, Imparato:2006}.
In particular, by setting $\ve = 0$, Eq.~(\ref{e.ldb0}) can be transformed as
\begin{equation}
\Pc_{\sigma_0} (x_0) P_0 ((\sigma_0, x_0) \to [\sigma, x]) = \Pc_{\sigma_N} (x_N) P_0 ((\sigma_N, x_N) \to [\sigma, x]).
\label{e.db0}
\end{equation}
This relation is termed the detailed-balance condition with respect to the canonical distribution given in Eq. (\ref{e.canonical}).
From this expression, it can be concluded that the canonical distribution is the stationary distribution in the system when Eq.~(\ref{e.db}) is satisfied.
However, when Eq. (6) is not satisfied, there is no stationary distribution with which the detailed-balance condition holds.

\subsubsection{Fluctuation-Dissipation Theorem}
We investigate the ensemble average of a quantity $A[\sigma, x]$, which is a functional of a path $[\sigma, x]$, when the initial distribution is the canonical distribution $\Pc_\sigma (x)$. The average is defined as
\begin{equation}
\bra A[\sigma, x] \ket_\ve \equiv \prod_{j = 0}^N \int_{-\infty}^\infty \d x_j \sum_{\sigma_j \in \{ 0, 1 \}} \Pc_{\sigma_0} (x_0) P_\ve ((\sigma_0, x_0) \to [\sigma, x]) A[\sigma, x].
\label{e.dav}
\end{equation}
By using Eq.~(\ref{e.ldb0}) and transforming variables from $[\sigma, x]$ to $\hat{[\sigma, x]}$, this expression can be rewritten as
\begin{eqnarray}
\bra A[\sigma, x] \ket_\ve &=& \prod_{j=0}^N \int_{-\infty}^\infty \d x_j \sum_{\sigma_j \in \{0, 1\}} \Pc_{\sigma_0} (x_0) \hat P_\ve ((\sigma_0, x_0) \to [\sigma, x]) \e^{-\beta\ve \sum_{k=0}^{N-1} \hfp{k}(x_{k+1} - x_k)} \hat A[\sigma, x] \nonumber \\
&=& \bra \hat A[\sigma, x] \ket_{\hat \ve} - \beta \ve \bra \sum_{j=0}^{N-1} \hfp{j}(x_{j+1}-x_j) \hat A[\sigma, x] \ket_0 + O(\ve^2),
\label{e.av}
\end{eqnarray}
where $\hat A[\sigma, x] \equiv A \hat{[\sigma, x]}$, and $\bra \cdots \ket_{\hat \ve}$ denotes the average over processes subject to the time-reversed perturbation $\hat \fp(t)$.

We now suppose that $N$ is an odd number and choose $A[\sigma, x] = (x_{(N+1)/2} - x_{(N-1)/2})/\Delta t$.
By setting $\ve = 0$ in Eq.~(\ref{e.av}), we immediately obtain
\begin{equation}
\bar v \equiv \bra \frac{x_{(N+1)/2} - x_{(N-1)/2}}{\Delta t} \ket_0 = 0.
\end{equation}
Furthermore, we choose the perturbation such that $\bfp{j} = \hfp{j} = \bfp{N-j-1}$. In this case, the two averages, $\bra \cdots \ket_\ve$ and $\bra \cdots \ket_{\hat \ve}$, coincide. Then, Eq.~(\ref{e.av}) yields
\begin{eqnarray}
\bra \frac{x_{(N+1)/2} - x_{(N-1)/2}}{\Delta t} \ket_\ve &=& \ve \sum_{j = 0}^{(N-3)/2} \beta C_{\frac{N-1}{2} - j} \bfp{j} \Delta t  + \ve \beta \frac{C_0}{2}  \bfp{(N-1)/2} \Delta t + O(\ve^2),
\label{e.response}
\end{eqnarray}
where
\begin{equation}
C_{j - k} \equiv \bra \frac{x_{j+1} - x_j}{\Delta t} \frac{x_{k+1} - x_k}{\Delta t} \ket_0
\label{e.corrd}
\end{equation}
is the velocity correlation function in the discretized form.
The fact that $C_{j - k}$ only depends on the difference $j - k$ is a consequence of the time-translational invariance of the system in a steady state including equilibrium.
$C_{j - k}$ converges to $C(t)$ defined in Eq.~(\ref{e.corr}) in the limit $\Delta t \to 0$ and $N \to \infty$ with $N \Delta t$ fixed (hereafter, referred to as ``the continuous-time limit'').

The definition of the response function [Eq.~(\ref{e.res})] can be discretized as 
\begin{eqnarray}
\lim_{\ve \to 0} \frac{1}{\ve} \bra \frac{x_{(N+1)/2} - x_{(N-1)/2}}{\Delta t} \ket_\ve &=& \sum_{j = 0}^{(N-1)/2} R \left(t_{(N-1)/2} +\frac{\Delta t}{2} - t_j \right) \fp(t_j) \Delta t \nonumber \\
&=& \sum_{j = 0}^{(N-1)/2} R_{\frac{N-1}{2}-j} \bfp{j} \Delta t + O(\Delta t),
\label{e.resd}
\end{eqnarray}
where $R_{j-k} \equiv R(t_j - t_k +\Delta t/2)$.
In the first line in Eq.~(\ref{e.resd}), we have employed Ito's scheme in discretizing the right-hand side of Eq.~(\ref{e.res}). In what follows, we will use this scheme, although the other schemes such as Stratonovich's scheme will lead to the same result if it is consistently used.

Comparison of Eqs.~(\ref{e.response}) and (\ref{e.resd}) yields
\begin{equation}
R_j = \left \{
\begin{array}{ll}
\beta C_j &\mathrm{for}\quad j \ge 1 \\
\beta C_{0}/2 &\mathrm{for}\quad j = 0 \\
0 & \mathrm{otherwise}
\end{array} \right.
\end{equation}
Therefore, we finally obtain the desired result Eq.~(\ref{e.tFDT}) in the continuous-time limit.

Next, we introduce the discrete Fourier transform of a series $\{ A_j \}$ as
\begin{equation}
\tilde A_\alpha \equiv \sum_{j = -(N-1)/2}^{(N-1)/2} \Delta t A_j \e^{2 \pi i \alpha j},
\end{equation}
which converges to the Fourier transform $\tilde A(\omega)$ with $\omega = 2 \pi \alpha/\Delta t$ in the continuous-time limit.
Using this notation, we calculate
\begin{eqnarray}
\tilde R_\alpha &=& \sum_{j = 0}^{(N-1)/2} \Delta t R_j \e^{2\pi i \alpha j} \nonumber \\
&=& \sum_{j = 1}^{(N-1)/2} \Delta t \beta C_j \e^{2\pi i \alpha j} + \Delta t \beta C_0 /2,
\end{eqnarray}
and
\begin{equation}
\tilde R_{-\alpha} = \sum_{j = -(N-1)/2}^{-1} \Delta t \beta C_j \e^{2\pi i \alpha j} + \Delta t \beta C_0 /2,
\end{equation}
where the symmetry property $C_j = C_{-j}$ has been used. Combining these expressions, it is found that
\begin{equation}
\beta \tilde C_{\alpha} = \tilde R_\alpha + \tilde R_{-\alpha}.
\end{equation}
Therefore, we obtain Eq.~(\ref{e.fFDT}) in the continuous-time limit.

\subsection{Proof of the main claim} \label{s.mainproof}

In this Section, we prove our main claim Eq.~(\ref{e.eq}) for the flashing ratchet model.
First, we consider the steady-state average of a quantity $A[\sigma, x] \equiv (x_{k+1} - x_k)/\Delta t$ by replacing $\Pc_{\sigma} (x)$ with $\Ps_{\sigma} (x)$ in Eq.~(\ref{e.dav}). In particular, the perturbation is chosen as $\bfp{j} = \delta_{jl}/\Delta t$.
Next, Eq.~(\ref{e.dav}) is rewritten in the following way
\begin{eqnarray}
\bra \frac{x_{k+1} - x_{k}}{\Delta t} \ket_\ve &=& \prod_{j_< = 0}^{l-1} \prod_{j_> = l+2}^{N} \int \d x_{j_<} \int \d x_{j_>} \sum_{\sigma_{j_<}, \sigma_{j_>}} \int \d x_l \int \d x_{l+1} \Ps_{\sigma_0} (x_0) \frac{x_{k+1} - x_k}{\Delta t} \nonumber \\
& & \times \Big[ P_\ve ((\sigma_0, x_0) \to [\sigma, x; \sigma_l = 0, \sigma_{l+1} = 0]) \nonumber \\
& & \quad \quad + P_\ve ((\sigma_0, x_0) \to [\sigma, x; \sigma_l = 0, \sigma_{l+1} = 1]) \nonumber \\
& & \quad \quad + P_\ve ((\sigma_0, x_0) \to [\sigma, x; \sigma_l = 1, \sigma_{l+1} = 0]) \nonumber \\
& & \quad \quad + P_\ve ((\sigma_0, x_0) \to [\sigma, x; \sigma_l = 1, \sigma_{l+1} = 1]) \Big].
\label{e.avf}
\end{eqnarray}
Then, by substituting Eq.~(\ref{e.path}) into this expression and differentiating it with respect to $\ve$, it is calculated 
\begin{eqnarray}
\lefteqn{\left. \pder{}{\ve} \bra \frac{x_{k+1} - x_k}{\Delta t} \ket_\ve \right|_{\ve \to 0} = \prod_{j_< = 0}^{l-1} \prod_{j_> = l+2}^{N} \int \d x_{j_<} \int \d x_{j_>} \sum_{\sigma_{j_<}, \sigma_{j_>}} \int \d x_l \int \d x_{l+1} \Ps_{\sigma_0} (x_0) \frac{x_{k+1} - x_k}{\Delta t} } \hspace{15mm} \nonumber \\ 
& & \times \bigg\{ \frac{1}{2 \gamma T} \left[ \gamma \frac{x_{l+1} - x_l}{\Delta t} + U'_0 (\bar x_l) \right] P_0 ((\sigma_0, x_0) \to [\sigma, x; \sigma_l = 0, \sigma_{l+1} = 0] ) \nonumber \\
& & \quad \quad + \frac{1}{2 \gamma T} \left[ \gamma \frac{x_{l+1} - x_l}{\Delta t} + U'_1 (\bar x_l) \right] P_0 ((\sigma_0, x_0) \to [\sigma, x; \sigma_l = 1, \sigma_{l+1} = 1] ) \bigg\} \nonumber \\
&=& \frac{1}{2\gamma T} \bra \left[ \gamma \frac{x_{l+1}- x_l}{\Delta t} + U'_{\sigma_l} (\bar x_l) \right] \frac{x_{k+1} - x_k}{\Delta t} \ket_0.
\end{eqnarray}
Comparing this expression with Eq.~(\ref{e.resd}) in which we set $\bfp{j} = \delta_{jl}/\Delta t$, we obtain the following relations
\begin{equation}
\frac{1}{2\gamma T} \left[ \bar v^2 + \gamma C_{k-l} + \bra U'_{\sigma_l} (\bar x_l) \frac{x_{k+1} - x_k}{\Delta t} \ket_0 \right] = \left\{ \begin{array}{ll} R_{k-l} & \quad \mathrm{for}\quad l \le k \\ 0 & \quad \mathrm{for} \quad l > k
\end{array} \right.
\label{e.hs}
\end{equation}
Eq.~(\ref{e.hs}) can be reexpressed as
\begin{eqnarray}
J_{j- k} &\equiv& - \frac{1}{2} \left[ \bra U'_{\sigma_j} (\bar x_j) \frac{x_{k+1} - x_k}{\Delta t} \ket_0 + \bra U'_{\sigma_k} (\bar x_k) \frac{x_{j+1} - x_j}{\Delta t} \ket_0 \right] \nonumber \\ 
&=& \gamma \left( \bar v^2 + C_{j -k} - \gamma T R_{j-k} \right)
\label{e.hs3}
\end{eqnarray}
for $k \le j$.
It is then found
\begin{eqnarray}
J_0 &=& - \bra U'_{\sigma_j} (\bar x_j) \frac{x_{j+1} - x_j}{\Delta t} \ket_0 \to  - \bra U_{\sigma(t)} (x(t)) \circ \dot x(t) \ket_0 = J,
\label{e.hsj}
\end{eqnarray}
in the continuous-time limit. In the same limit, the right-hand side of Eq.~(\ref{e.hs3}) converges as
\begin{equation}
\gamma \left( \bar v^2 + C_0 - \gamma T R_0 \right) \to \gamma \left\{ \bar v^2 + \int_{-\infty}^\infty \left[ \tilde C(\omega) - T \left( \tilde R(\omega) + \tilde R(-\omega) \right) \right] \frac{\d \omega}{2\pi} \right \},
\label{e.hsv}
\end{equation}
where Fourier's integration theorem
\begin{equation}
\lim_{\Delta t \to 0} \frac{R(-\Delta t/2) + R(\Delta t/2)}{2} = \int_{-\infty}^\infty \frac{\tilde R(\omega) + \tilde R(-\omega)}{2} \frac{\d \omega}{2\pi}
\end{equation}
has been used.
By combining these expressions, we finally obtain the desired result in Eq.~(\ref{e.eq}).

\section{Discussions and Conclusion} \label{s.diss}
\subsection{Applications to experiments} \label{s.appl}
In this Section, we discuss the possibility and significance of applying our result to experimental studies on molecular motors.

As explained in Sec.~\ref{s.main}, the energy input to the system per unit time is obtained using Eq.~(\ref{e.eq}) by determining the mean velocity, $\bar v$, the power spectrum density, $\tilde C(\omega)$, and the susceptibility, $\tilde R(\omega)$, through measurements.
This procedure requires no more detailed knowledge on the system such as the potential profile.
Therefore, if this method can be used in experimental studies of molecular motors, it will be a great advantage, since experimental determination of energy input rate is technically quite difficult.
However, there are several restrictions in order to warrant the relevance of this method to a real molecular motor system.

First, a motor system to be considered has to be reasonably well described with some kind of Langevin model including a Brownian motor model, since the present theory is constructed on the framework of a Langevin equation. This means that the position of the center of mass of the motor, $x(t)$, has to be a considerably slow variable compared with the other degrees of freedom in the system.
If there is another degree of freedom, $y(t)$, that has a comparable timescale with that of $x(t)$, we have to include $y(t)$ in the system of Langevin equations to be considered. In this case, the quantity $J$ given in Eq.~(\ref{e.eq}) provides the information how much energy is dissipated through the degree of freedom $x(t)$ per unit time. The total energy input will be the sum of all the dissipated energy through degrees of freedom $x(t)$ and $y(t)$.

This condition might be satisfied in some cases. In particular, since an $\mathrm{F}_1$ motor has a single degree of freedom for rotation, it is plausible that this molecule can be described using a Langevin equation where the slow variable is a rotational angle of the rotational unit ($\gamma$ subunit). A kind of kinesin that has only one head part, termed KIF1A \cite{Okada:1999}, is another example, since the motion of this molecule could be described with a Langevin equation where the slow variable is the position of the center of mass.
Furthermore, a conventional kinesin that has two heads can also be modeled by a Langevin equation with a single variable, which is the position of the center of mass, because this molecule uses two heads alternatively and thus a single variable is enough to represent the state of the molecule.

The most important point here is that the validity of these conjectures can be experimentally determined through examination of Eq.~(\ref{e.eq}). Because we can roughly estimate the free energy consumed by a single molecule per unit time by a macroscopic biochemical experiments, we can determine whether Eq.~(\ref{e.eq}) is satisfied in a molecule under consideration by adopting this macroscopic estimate of the energy input rate as $J$.
If Eq.~(\ref{e.eq}) is verified in experiments, this implies that our assignment of slow variables was correct. On the contrary, if it is denied experimentally, this implies that there are other slow variables and our knowledge on the system is insufficient.
In this manner, experimental examination of Eq.~(\ref{e.eq}) can be used as a ``check sum'' to single out the slow variables that is involved in the energy transduction.

Second, in order to correctly determine the right-hand side of Eq.~(\ref{e.eq}), the time and spatial resolutions in experiments have to be sufficiently good, although very good time resolution is often difficult to be achieved.
However, by noticing that the integrant of the second term on the right-hand side of Eq.~(\ref{e.eq}) vanishes for high frequencies as seen in Fig.~\ref{f.FDT} (b), it is enough to determine $\tilde C(\omega)$ and $\tilde R(\omega)$ up to a finite frequency $\omega^*$ (the decay rate).
From the investigation of the flashing ratchet model, it is conjectured that $2\pi/\omega^*$ is close to the time needed for a motor to diffuse around a single potential well.
Considering the current standard of the single-molecule techniques, this requirement seems to be within the scope.

\subsection{Example: KIF1A}
Here, we attempt to apply the present theory to KIF1A as an illustration.
Unfortunately, the experimental data currently available in the literature of KIF1A is insufficient to determine $\tilde C(\omega)$ and $\tilde R(\omega)$ in a sufficiently large frequency domain.
Thus, we cannot verify the validity of Eq.~(\ref{e.eq}) for this molecule at present.

Instead, we here determine the decay rate $\omega^*$ of KIF1A by assuming that the KIF1A can be modeled by the flashing ratchet.
For this purpose, we employ an approximation for the right-hand side of Eq.~(\ref{e.eq}).
In the calculation of the FRR violation shown in Fig.~\ref{f.FDT} (b), it is found that the function $\tilde C(\omega) - 2 \kb T \tilde R^{+} (\omega)$ is well approximated by a Lorentzian, $[\tilde C(0) - 2 \kb T \tilde R^{+} (0)] (\omega^*)^2 / [ \omega^2 + (\omega^*)^2 ]$.
Thus, using this approximation in the second term on the right-hand side of Eq.~(\ref{e.eq}), we calculate
\begin{equation}
\int_{-\infty}^\infty \left[\tilde C(\omega) - 2 \kb T \tilde R^{+} (\omega) \right] \frac{\d \omega}{2\pi} \approx (D - \mu \kb T) \omega^*,
\end{equation}
by use of the relations $2D = \tilde C(0)$ and $\mu = \tilde R^{+} (0)$.

We then adopted the following values from Refs.~\cite{Okada:1999, Okada:2000, Okada:2003}: $\bar v = 200$ (nm/s), $D = 4 \times 10^4$ ($\mathrm{nm^2}$/s), $\mu = 1.3 \times 10^3$ (nm/s$\cdot$pN) in the presence of 2 mM ATP at $T = 300$ (K), and $\gamma = 1 \times 10^{-4}$ (pN$\cdot$s/nm). The value of $\gamma$ is determined as follows; when a KIF1A molecule is traped in a ADP-binding state, it freely diffuses on a microtubule with a diffusion coefficient $D_\mathrm{ADP}$ without energy consumption. We determined the value of $\gamma$ through Einstein's relation as $\gamma = \kb T/D_\mathrm{ADP}$ by assuming that this is a free diffusion.

As the value of $J$, we adopt the rate of free energy consumption through ATP hydrolysis reactions.
In physiological condition, it is believed that a single hydrolysis reaction of ATP molecule produces the free energy of $\Delta G = 80$ (pN$\cdot$nm). A KIF1A molecule hydrolyses ATP molecules at the rate $\nu = 1 \times 10^2$ ($\mathrm{s}^{-1}$) in the presence of 2 mM ATP. Therefore, the energy consumption by a single motor per unit time is estimated as $\nu \Delta G = 8 \times 10^3$ (pN$\cdot$nm/s).

We then determine the value of $\omega^*$ through
\begin{equation}
\omega^* = \frac{J - \gamma \bar v^2}{\gamma (D - \mu \kb T) }.
\end{equation}
Calculation using above mentioned experimental data yields $\omega^* / 2\pi = 4 \times 10^2$ ($\mathrm{s}^{-1}$).
This value of the decay rate is not much different from the inverse of the cycle time of the ATP hydrolysis reactions and the dwell time of a KIF1A, it should be possible to determine the relevance of this value by use of the single-molecule experimental techniques.
If the validity of the above obtained value of $\omega^*$ is experimentally verified, this implies that the flashing ratchet model is quantitatively relevant for the KIF1A.

\subsection{Concluding Remark} \label{s.conc}

In conclusion, a new equality relating the rates of energy input and dissipation to the extent of the FRR violation, which has been obtained by the authors recently, has been explained in detail.
This relation enables us to investigate the energetics of a Langevin system on the basis of experimentally measurable quantities alone.
With regard to the flashing ratchet model, the FRR and this new relation is numerically demonstrated and mathematically proved.
It was also explained that this new relation can be used in experimental studies of molecular motors in order to single out the slow variables related to the energy transduction. An attempt to apply this theory to the KIF1A has been presented and the value of the decay rate has been predicted.

We believe that to test this result for various types of molecular motors is important to elucidate the mechanism of their mechano-chemical energy transduction. Our method provides a novel approach in the study of molecular motors since it enables determination of energy flows at the single-molecule level.
We also hope that by extending the present approach, we can establish a methodology to extract information of nonequilibrium systems including biological molecular machinery by external measurements.

\vspace{12pt}

The authors acknowledge H. Noji, R. Iino, K. Tabata, and K. Hayashi for discussions on several issues in this paper.
The authors are also grateful to G. C. Paquette for his advices to improve the presentation of the paper.
This study was supported by a grant from the Ministry of Education, Science, Sports and Culture of Japan, No.~16540337 and Research Fellowships for Young Scientists from the Japan Society for the Promotion of Science, No.~05494.

\end{document}